\documentclass[aps,superscriptaddress,twocolumn,showpacs,preprintnumbers,amsmath,amssymb]{revtex4}
\usepackage[cp1251]{inputenc}
\usepackage{amssymb,amsmath,textcomp, wasysym}
\usepackage[mathscr]{eucal}
\usepackage{graphicx}
\usepackage{longtable}
\usepackage[dvips]{hyperref} 
\begin{document}
\title{Rabi Wave Packets and Peculiarities of Raman Scattering  in   Carbon  Nanotubes, Produced by High Energy Ion Beam Modification of Diamond Single Crystals}
\author{Dmitry Yearchuck (a),  Alla Dovlatova (b)\\
\textit{(a) - Minsk State Higher Aviation College, Uborevich Str., 77, Minsk, 220096, RB; yearchuck@gmail.com,  \\ (b) - M.V.Lomonosov Moscow State University, Moscow, 119899}}
\date{\today}
\begin{abstract}  QED-model for  multichain coupled qubit system, proposed in \cite{Part1}, was confirmed by Raman scattering studies of quasi-1D carbon zigzag-shaped nanotubes (CZSNTs), produced by high energy ion beam modification of natural diamond single crystals.  Multichain coupled qubit system represents itself Su-Schriffer-Heeger $\sigma$-polaron lattice, formed in CZSNTs plus  quantized external electromagnetic (EM) field. New quantum optics phenomenon - Rabi waves, predicted in \cite{Slepyan_Yerchak} has experimentally been  identified for the first time. It is shown, that Raman spectra in  quasi-1D CZSNTs are quite different in comparison with well known Raman spectra in 2D those ones. They characterized by semiclassical consideration by the only  one vibronic mode of Su-Schriffer-Heeger $\sigma$-polaron lattice instead of longitudinal
and transverse optical phonon    $G^+$ and $G^-$modes and
the out-of-plane radial breathing mode, which are observed in Raman spectra of  2D single wall nanotubes. It is consequence of 2D - 1D transition in all physical properties of nanotubes. It is shown, that  strong electron-photon coupling takes place in CZSNTs by interaction with EM-field and quantum nature of EM-field has to be taken into account. It has been done for the first time in stationary spectroscopy at all. All optical spectra, in particular, Raman spectra are registered by usual stationary measurement technique  in nonequilibrium conditions, which are the consequence of Rabi wave packets' formation. It leads in its turn  to appearance of additional lines, corresponding to revival part of inversion dependence of joint EM-field + matter system in frequency representation.  
\end{abstract}
\pacs{42.50.Ct, 61.46.Fg, 73.22.–f, 78.67.Ch, 77.90.+k, 76.50.+g}
\maketitle 
\section{Introduction}
Quantum electrodynamics (QED) takes on more and more significance for its practical application and it, in fact, becomes to be working instrument in spectroscopy studies and industrial spectroscopy control. 
 QED-model for  multichain coupled qubit system was proposed in \cite{Part1}. It     is generalization of the  model described in \cite{Slepyan_Yerchak}, generalizing, in its turn, Tavis-Cummings model \cite{Tavis} by taking into account the 1D-coupling between qubits. The most substantial result in \cite{Slepyan_Yerchak} is the prediction of new quantum optics phenomenon - Rabi waves and Rabi Wave packets' formation.
  It is substantial, that in the model, proposed in \cite{Part1} the interaction of quantized EM-field with multichain qubit system  is considered by taking into account both the intrachain and interchain qubit coupling. Tne model is presented with an example of perfect carbon zigzag-shaped carbon nanotubes (CZSNTs). CZSNTs  can be considered  to be the set of $n$ carbon backbones of trans-polyacetylene (t-PA) chains,  which are connected between themselves. Given $n$-chain  set can be  considered to be a single whole, which  holds the quasi-one-dimensionality of a single chain. It seems to be correct for perfect CZSNTs, if their diameter is $\leq$ 1 nm (see Sec.III). At the same time it is well known, that free standing nanotubes were considered theoretically   to be $2D$-strutures and two-dimensional lattice structure of a single wall carbon nanotube (SWNT)
is specified uniquely by the chirality defined by two integers
$(n,m)$ \cite{Dresselhaus}, \cite{Saito}.  Two in-plane $G$ point longitudinal and transverse optical phonon $(LO$ and $TO)$ modes \cite{Reich} and
the out-of-plane radial breathing mode $(RBM)$ \cite{Dresselhaus M.S} are observed in the Raman spectra of  SWNTs.
The $LO$ and $TO$ phonon modes at
the $G$ point in the two-dimensional Brillouin zone
are degenerate in graphite and graphene, however they split in SWNTs into two peaks,
denoted by $G^+$ and $G^-$peaks, respectively, \cite{G.Dresselhaus} because of the
curvature effect. The agreement of experimental Raman studies of carbon nanotubes (NTs) with diameter $ \apprge 1 nm$ with $2D$ SWNT-theory unambiguously indicates, that given NTs, produced  by CVD-methods and like them are really $2D$ systems.  At the same time the narrow  NTs with diameter $ < 1 nm$ cannot be considered strongly speaking to be  $2D$-systems, they are quasi-$1D$ systems, at that 2D-1D transition takes place, theoretical explanation of which has been done for the first time in \cite{Part1}. Earlier experimental results for 2D-1D transition, obtained by electron spin resonance (ESR) method in quasi-1D CZSNTs, produced by high energy ion beam modification (HEIBM) of natural diamond single crystals will be summarized in presented work   and new experimental results obtained by Raman scattering method will be presented.

The aim of given work is also to represent the experimental evidence for multichain QED-model, proposed in \cite{Part1} and to confirm the theoretical prediction of  Rabi wave phenomenon in \cite{Slepyan_Yerchak}. The new phenomenon predicted is quantum coherent effect. It means, that along with requirement of quasionedimensionality the  requirement of structural perfectness arises. To observe quantum optical coherent effects on NTs, the ensemble of NTs has to be homogeneous. It means, that any dispersion in axis direction, chirality, length and especially in diameter both for single NT along its axis and between different NTs in ensemble has to be absent, axial symmetry has to be also retained, that is, there are additional requirements in comparison with, for example, t-PA technology. The CVD-technology of NTs production and many similar to its ones seem to be not satisfying to above-listed requirements at present. It means, that experimental results and their theoretical treatment will be different in both the cases, that really takes place (see for more details Sec.III and Sec.IV). The situation seems to be analogous to some extent to the solid state physics of the same substance in single crystal and amorphous forms. The technology, based on HEIBM,  satisfy given requirements.  

The paper is organized in the following way. In Section II, experimental technique and Raman scattering results are described. In Section III, the comment and some development of Su, Schrieffer, Heeger (SSH) model for organic conductors
are given. In Section IV, experimental results are discussed. In Section V, the conclusions  are presented.

 \section{Experimental Technique and Results}

Samples of type IIa natural diamond, implanted by high energy ions of copper $(63$ $MeV$, $5\times{10^{14}}$ $cm^{-2}$) and boron $(13,6$ $MeV)$ have been studied. Ion implantation was performed along $\left\langle{111}\right\rangle$ crystal direction. Raman scattering  (RS) spectra were registered in backscattering geometry. Laser excitation wave length was 488 $nm$, rectangular slit $350{\times}350 (\mu m)^2$ was used, scan velocity was 100 $cm^{-1}$ pro minute.
 
The spectra observed are presented in Figures 1 to 3. 
All the spectra  are $180 \textdegree$ out of phase, that is dip positions correspond to maxima of the signals.   The spectrum, presented in Figure 1,  is characteristic of the only ion beam modified region (IBMR) of the sample, since the RS-line near 1332 $cm^{-1}$, characteristic for diamond single crystals, is absent in the spectrum. The RS-lines with peak positions $656.8{\pm}0.2$ $cm^{-1}$, $1215{\pm}1$ $cm^{-1}$, $1779.5{\pm}1$ $cm^{-1}$ and  $2022.3{\pm}0.5$ $cm^{-1}$ correspond to optical centers in IBMR, which are active   by laser excitation transversely to sample surface from implanted side. At the same time, by laser excitation of the same sample from opposite unimplanted side the qualitatively other picture is observed, see Figure 2. Firstly, the line,  characteristic for diamond single crystals, is now  presented in the spectrum (the whole amplitude of given line is not shown in Figure 2). It is interesting, that its frequency value (1328.7 $cm^{-1}$) is slightly shifted from usually observed value near 1332 $cm^{-1}$. It  indicates on renormalization of optical phonon by Cu-implantation in the whole diamond matrix.

 Secondly, RS-lines with peak positions  at 354.6, 641.8, 977.1 (${\pm}1$ $cm^{-1}$), 1274.1 ${\pm}2$ $cm^{-1}$  and more weak pronounced lines at 1569 ${\pm}3$ $cm^{-1}$, 1757${\pm}5$ $cm^{-1}$ were observed. Moreover, all the lines listed are superimposed now with very broad (its linewidth  is 1720${\pm}20$ $cm^{-1}$) asymmetric line with peak position $1160{\pm}10$$cm^{-1}$. 

The spectrum of boron implanted sample, Figure 3, was measured in the range (1000 - 2100) $cm^{-1}$. It is seen, that the spectra of boron and copper implanted samples are qualitatively similar in given spectral range. However numerical values  - 1212.3${\pm}1$,  1772.5${\pm}1$, 2011${\pm}0.5$$cm^{-1}$ - of the peak positions are slightly different, they are shifted to low frequency region. It is interesting, that there is regularity in peak position shift, which is increasing  with frequency increase, and it is equal to 2.8, 7, 11.3 $cm^{-1}$ correspondingly.  

\begin{figure}
\includegraphics[width=0.5\textwidth]{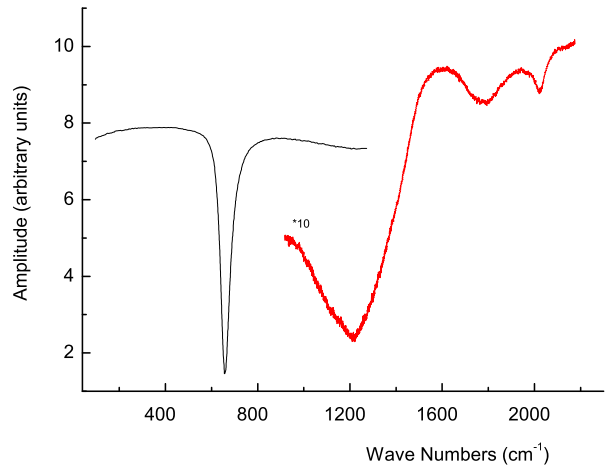}
\caption[Spectral distribution of Raman scattering  intensity in
diamond single crystal, implanted by high energy copper ions, the excitation is from implanted side of the sample.]
{\label{Figure1} Spectral distribution of Raman scattering  intensity in
diamond single crystal, implanted by high energy copper ions, the excitation is from implanted side of the sample.}
\end{figure}
 \begin{figure}
\includegraphics[width=0.5\textwidth]{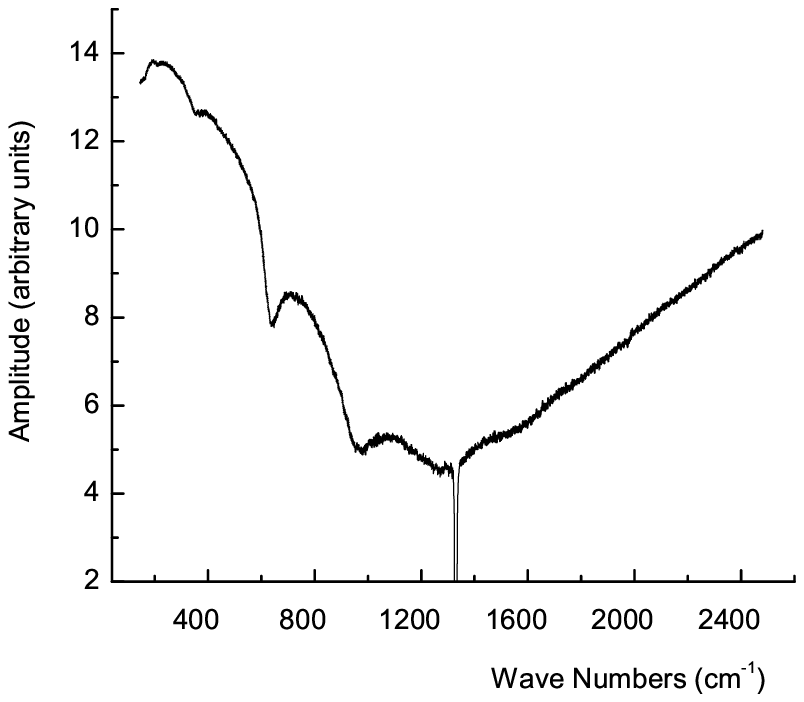}
\caption[Spectral distribution of Raman scattering  intensity in
diamond single crystal, implanted by high energy copper ions, the excitation is from unimplanted side of the sample.]
{\label{Figure2} Spectral distribution of Raman scattering  intensity in
diamond single crystal, implanted by high energy copper ions, the excitation is from unimplanted side of the sample.}
\end{figure}
\begin{figure}
\includegraphics[width=0.5\textwidth]{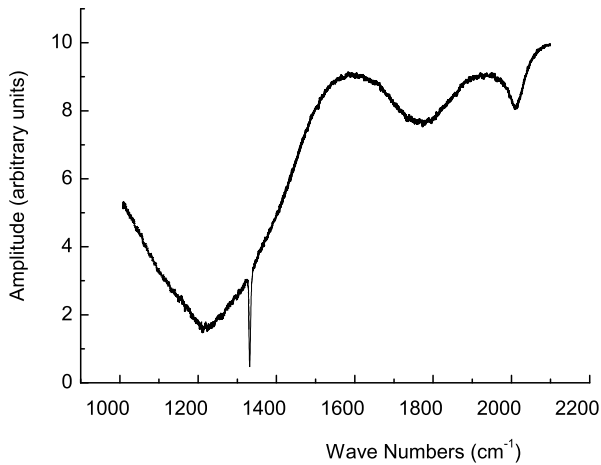}
\caption[Spectral distribution of Raman scattering  intensity in
diamond single crystal, implanted by high energy boron ions, the excitation is from implanted side of the sample.]
{\label{Figure3} Spectral distribution of Raman scattering  intensity in
diamond single crystal, implanted by high energy boron ions, the excitation is from implanted side of the sample.}
\end{figure}
 It is strong indication, that all three lines belong to the same optical system in both the samples studied. The  line with peak position 1331.95 ${\pm} 0.1$$cm^{-1}$ was also presenting. Its frequency value coincides with the value of optical phonon peak in conventional natural diamonds. The presence of given RS-line by excitation from implanted side can be determined by two factors. Firstly, the effective thickness of IBMR is substantially less, it is $\simeq{1.4}\mu m$, in comparison with Cu-implanted sample ($\simeq{7}\mu m$) \cite{Erchak}, consequently, laser excitation can reach an unimplanted region. It can also be suggested, that the modification is not entire in near surface region.  Given results seem to be the first results in application of RS-spectroscopy to study of IBMR in diamonds.  Let us also to pay attention, that the length of CZSNTs is determined by the implantation with the ion energy $\sim 1 Mev$ per nucleon by the thickness of IBMR, that is it gets in usual for free standing NTs micron-th range. It follows from ESR experiments on layer-by-layer removal of IBMR, see \cite{Ertchak} and the references therein.

Comparing RS-spectra, represented in Figures 1 to 3 with well known spectra of NTs produced by CVD-methods it is seen, that difference between the optical characteristics of the NTs, produced by HEIBM of diamond single crystals and  between the NTs, produced by other methods, including methods of ultra-small NTs' production in zeolite matrix, is  greatly, while there is the essential similarity of ESR-data with those ones for ultra-small NTs, embedded in zeolite matrix (see Sections III, IV).   

On the other hand,
comparing the figures 1 and 2  between themselves, we can straight away conclude, that RS-lines observed cannot be attributed to usual atomic or molecular vibrations including the local vibration modes like to well known G-bands and  RBM, observed in  NTs, produced by CVD-methods, since the RS-spectra, corresponding to usual local vibrations cannot be dependent on the change of  propagation direction of exciting light into opposite.

Therefore 2D-theory, developed for explanation of optical properties of 2D-NTs cannot be applicable for perfect NTs of small diameter. It is strong argument for  application of the QED-model, proposed in \cite {Part1} for correct description of  the objects studied.

Since the theoretical  model of small quasi-1D CZSNTs \cite {Part1} is based partly on SSH-model of 1D organic conjugated conductors let us consider SSH-model more carefully. It will be shown, that it needs in some corrections and in development. 

 \section{Comment to SSH-Model for Organic Conductors and Some its Development}

 It is substantial, that  the physical properties of  t-PA chains are studied very well both theoretically and experimentally. Let us touch on given subject in more detail.   Su, Schrieffer, Heeger \cite{Su_Schrieffer_Heeger_1979}, \cite{Su_Schrieffer_Heeger_1980} have found the most simple way to describe mathematically  the chain of t-PA by considering it to be Fermi liquid. It is now well known SSH-model. The most substantial suggestion in SSH-model is concerned the number of degrees of freedom. Su, Schrieffer, Heeger  suggested, that the only dimerization coordinate $u_n$ of the $n$-th $CH$-group, 
$n = \overline{1,N}$ along chain molecular-symmetry axis $x$ is essential for determination of main physical properties of t-PA. Other five degrees of freedom were not taken into consideration. Nevertheless, the model has obtained magnificent experimental confirmation, that requires additional argumentation.   
At the same time it will be shown in the next subsection, that SSH-model itself needs in some corrections, which being to be of the principle lead nevertheless qualitatively to analogous conclusions, concerning physical properties of the system studied.

 \subsection{Corrections to SSH-Model}

We have repeated the calculation, presented  in \cite{Su_Schrieffer_Heeger_1979}, \cite{Su_Schrieffer_Heeger_1980}. It has been found, that there is the second solution, which leads to independent branch of quasiparticles. Let us show it. We preserve all designations for quantities from \cite{Su_Schrieffer_Heeger_1980}.

  The expression for Hamiltonian in reduced zone has been obtained in the following form
\begin{equation}
\label{Eq1}
\begin{split}
&\hat{H}(u) = \sum_{k}\sum_{s}[\varepsilon_k ({\alpha}^2_k - {\beta}^2_k) + 2 {\alpha}_k{\beta}_k \Delta_k](\hat{n}^c_{ks} - \hat{n}^v_{ks})\\
&+ 2NKu^2\hat{e},
\end{split}
\end{equation}
where ${\alpha}_k$ and ${\beta}_k$  are determined by
\begin{equation}
\label{Eq2}
\alpha_k = \sqrt {\frac{1 \mp \frac{\varepsilon_k}{E_k}}{2}},
{\beta}_k = \sqrt {\frac{1 \pm \frac{\varepsilon_k}{E_k}}{2}}.
\end{equation}
$E_k$ is determined by expression
\begin{equation}
\label{Eq3}
E_k = \sqrt{\varepsilon^2_k  + \Delta^2_k},
\end{equation}
in which
\begin{equation}
\label{Eq4}
\varepsilon_k = 2t_0\cos{ka},
\Delta_k = 4\alpha u \sin{ka}
\end{equation}
and $\hat{n}^c_{ks}$, $\hat{n}^v_{ks}$ are occupation number operators in $C$- and $V$-bands correspondingly, $\hat{e}$ is unit operator.
The known SSH-solution corresponds to choice of lower signs in relations (\ref{Eq2}). The choice of upper signs gives independent solution. Therefore, for the energy of quasiparticles $E^{c[u]}_k$, $E^{v[u]}_k$ in $C$- and $V$-bands, corresponding to upper-sign-solution we obtain
\begin{equation}
\label{Eq5}
\begin{split}
&E^{c[u]}_k = \frac{\delta E^{[u]}(u)}{\delta{n}^c_{ks}} = \frac{\Delta^2_k - \varepsilon^2_k}{E_k},\\
&E^{v[u]}_k = 
\frac{\delta E^{[u]}(u)}{\delta{n}^v_{ks}} =  -(\frac{\Delta^2_k - \varepsilon^2_k}{E_k}),
\end{split}
\end{equation}
We see, that it differs from the energy of  quasiparticles $E^{c[l]}_k$, $E^{v[l]}_k$ in $C$- and $V$-bands, corresponding to lower-sign-solution (SSH-solution), which is 
\begin{equation}
\label{Eq6}
E^{c[l]}_k = \frac{\delta E^{[l]}(u)}{\delta{n}^c_{ks}} = E_k, E^{v[l]}_k = \frac{\delta E^{[l]}(u)}{\delta{n}^v_{ks}} = -E_k.
\end{equation} 
$E^{[u]}(u)$ and  $E^{[u]}(u)$ in (\ref{Eq5}) and (\ref{Eq6}) are eigenvalues of operator $\hat{H}(u)$, which correspond to upper and lower signs in (\ref{Eq2}), 
 ${n}^c_{ks}$, ${n}^v_{ks}$ are eigenvalues of operators of particle numbers in $C$-band and $V$-band correspondingly. The coefficients  ${\alpha}_k$ and ${\beta}_k$  were found in \cite{Su_Schrieffer_Heeger_1980}  from the conditions of energy minimum. However, the only necessary conditions were used. At the same time the  sufficient conditions for the minimum are substantial in given case, they change the role of both solutions. Really, since  ${\alpha}_k$ and ${\beta}_k$ are coupled by the condition
\begin{equation}
\label{Eq7}
{\alpha}^2_k + {\beta}^2_k = 1,
\end{equation} 
 we have conditional extremum and  sufficient conditions for the minimum
 can be  obtained by standard study.  The second differential of  the energy to be the function of  three variables  ${\alpha}_k$,  ${\beta}_k$ and $\lambda$, where 
 $\lambda$ is coefficient in Lagrange function for conditional extremum, has to be positively defined quadratic form. From the condition of positiveness of three principal minors of quadratic form coefficients we obtain  the following three sufficient conditions for the energy minimum

\paragraph{The first condition}

\begin{equation}
\label{Eq8}
\begin{split}
&\{ \varepsilon_k (1 - \frac{\varepsilon_k}{E_k}) < \frac{\Delta^2_k}{E_k} | 
({n}^c_{ks} - {n}^v_{ks}) < 0\}, \\
&\{\varepsilon_k (1 - \frac{\varepsilon_k}{E_k} > \frac{\Delta^2_k}{E_k} | ({n}^c_{ks} - {n}^v_{ks}) > 0 \}
\end{split}
\end{equation}
for the SSH-solution and

\begin{equation}
\label{Eq9}
\begin{split}
&\{ \varepsilon_k (1 + \frac{\varepsilon_k}{E_k} < \frac{\Delta^2_k}{E_k} | ({n}^c_{ks} - {n}^v_{ks}) < 0 \},\\
&\{\varepsilon_k (1 + \frac{\varepsilon_k}{E_k} > \frac{\Delta^2_k}{E_k} | ({n}^c_{ks} - {n}^v_{ks}) > 0 \}
\end{split}
\end{equation}
 for the additional solution. It is seen, that the first condition is realizable for the quasiparticles of both the kinds, at that in equilibrium $({n}^c_{ks} - {n}^v_{ks} < 0)$  and in  nonequilibrium $(n^c_{ks} - {n}^v_{ks} > 0$ conditions.  

\paragraph{The second condition}

The second condition is the same for both the solutions and it is
\begin{equation}
\label{Eq10}
 (\frac{\varepsilon^2_k}{E_k} - 2\frac{\Delta^2_k}{E_k})^2 - E^2_k  + \frac{3}{4} \Delta^2_k > 0 
\end{equation}

\paragraph{The third condition}

For the SSH-solution we have
\begin{equation}\label{Eq11}
(3\frac{\Delta^2_k}{E_k} + 4\frac{\varepsilon^2_k}{E_k})({n}^c_{ks} - {n}^v_{ks}) > 0. 
\end{equation}

It  means,  that   SSH-solution is unapplicable for description of standard processes, passing near equilibrium state by any parameters. The quasiparticles, described by   SSH-solution, can be created the only in strongly nonequilibrium state with inverse                                                                                                   
population of the levels in $C$- and $V$-bands. At the same time for the solution, which corresponds to upper signs in (\ref{Eq2}), we obtain 
\begin{equation}
\label{Eq12}
(3\frac{\Delta^2_k}{E_k} - 4\frac{\varepsilon^2_k}{E_k})({n}^c_{ks} - {n}^v_{ks}) > 0, 
\end{equation}
which is realizable both in near equilibrium and in strongly nonequilibrium states of the systems by corresponding choose of parameters.

Let us consider the continuum limit for the ground state of the $t$-PA chain with quasiparticles of given branch. Taking into account, that in ground state ${n}^c_{ks} = 0$, ${n}^v_{ks} = 1$ we have
\begin{equation}
\label{Eq13}
E^{[u]}_0(u) = - \frac{2Na}{\pi}\int\limits_0^{\frac{\pi}{2a}} \frac{\Delta^2_k - 
\varepsilon^2_k}{ \sqrt{\Delta^2_k + 
\varepsilon^2_k}}dk + 2NKu^2,
\end{equation}
then, calculating the integral, we obtain
\begin{equation}
\label{Eq14}
\begin{split}
&E^{[u]}_0(u) =  \frac{4Nt_0}{\pi}\{F(\frac{\pi}{2}, 1 - z^2) + \\
&\frac{1 + z^2}{1 - z^2}[E(\frac{\pi}{2}, 1 - z^2) - F(\frac{\pi}{2}, 1 - z^2)]\} + 2NKu^2, 
\end{split}
\end{equation}
where
 $F(\frac{\pi}{2}, 1 - z^2)$ is the complete elliptic integral of the first kind, 
$E(\frac{\pi}{2}, 1 - z^2)$ is the complete elliptic integral of the second kind,
$z^2 = \frac{2\alpha u}{t_0}$.
Approximation of ({\ref{Eq14}}) at $z \ll 1$ gives
\begin{equation}
\label{Eq15}
\begin{split}
&E^{[u]}_0(u) = N \{\frac{4t_0}{\pi} - \frac{6}{\pi}\ln\frac{2t_0}{\alpha u} \frac{4 \alpha^2 u^2}{t_0} + \\
&\frac{28 \alpha^2 u^2}{\pi t_0} + ...\} + 2NKu^2.
\end{split}
\end{equation}
It is seen from (\ref{Eq15}), that the energy of quasiparticles, described by   solution, which corresponds to upper signs in (\ref{Eq2}) has the form of Coleman-Weinberg potential with two minima at the values of dimerization coordinate $u_0$ and $-u_0$ like to energy of quasiparticles, described by   SSH-solution \cite{Su_Schrieffer_Heeger_1980}. 

Therefore, all qualitative conclusions of the model proposed in \cite{Su_Schrieffer_Heeger_1980} are holding, however for the quasiparticles, corresponding to the second-branch-solution.

 \subsection{Slater Principle and SSH-Model}

In \cite{Part1} was  suggested, that  success of SSH-model is the consequence of some general principle and it was shown, that    given general principle is really exists. Main idea was proposed by Slater at the earliest stage of quantum physics era already in 1924, that is before creation of quantum mechanics and quantum electrodynamics. It is - "Any atom may in fact be supposed to communicate with other atoms all the time it is in stationary state, by means of virtual field of radiation, originating from oscillators having the frequencies of possible quantum transitions..." \cite{Slater}. The development of given idea is based on the results of the work \cite {Dovlatova}. It has been found  in \cite {Dovlatova}, that Coulomb field in 1D-systems or 2D-systems can be quantized, that is, it has the character of radiation field and it can exist without the sources, which have created given field. Consequently, Coulomb field can be considered to be "virtual" field  in Slater principle and it can be applied to both t-PA and to quasi-1D-NTs.  It produces in t-PA the preferential direction in atom  communication the only along chain axis (to be consequence of quasi-one-dimensionality). It is reasonable to suggest, that  given direction remains to be also preferential by interaction with external EM-field, then the explanation of the success of SSH-model, taking into consideration the only one degree of freedom of $n$-th $CH$-group instead of six ones becomes to be natural.   It determines also  the applicability of SSH-model in the part concerning active degrees of freedom by interaction with external EM-field to  quasi-one-dimensional  CZSNTs, while the interaction between the chains, which produce CZSNTs is  strong (see further). 
 
Moreover Slater principle can be applied to  2D-SWNTs. However the physical consequences of Slater principle application are quite different for quasi-1D and 2D systems. In quasi-1D-systems, that is in t-PA and in quasi-1D SWNTs Coulomb field can be considered to be "virtual" field with propagation direction the only along  t-PA chain and NT-axis correspondingly.  In other words it produces preferential direction in atom  communication the only in one direction (to be consequence of quasi-one-dimensionality), and given direction remains to be preferential by interaction with external EM-field, that explains qualitatively the success of SSH-model in the sense that degrees of freedom, realized by bonds, projections of which  are not coinciding  with chain molecular-symmetry axis direction, can really be not taken into consideration for experiments with the participation of external EM-field. The consequence of given conclusion is the following. Quasi-1D CZSNT can be represented by the model consisting of the set of $n$ interacting between themseves equivalent carbon backbones of t-PA chains, while in real quasi-1D  CZSNTs  the adjacent chains are representing the mirror image of each other (relatively the corresponding radial plane). Given detail of SSH-model is used further in the model of quasi-1D  CZSNTs.  At the same time, there are existing in 2D-systems including 2D-SWNTs    two preferential directions of  Coulomb "virtual" field  propagation. It means, that for 2D-SWNTs  all degrees of freedom, which are relevant to the bonds in rolled graphene sheet will be essential by interaction with external EM-field and the simplification, which is fruitfully used in 1D-SSH model becomes to be incorrect for 2D-SWNTs in full accordance with existing 2D-SWNTs theory and Raman scattering experiments, see for instance \cite{Dresselhaus},  \cite{Reich}, \cite{G.Dresselhaus}, \cite{Dresselhaus M.S}.  

 Experimental confirmation for applicability of Slater principle to quasi-1D-CZSNTs and for the model of quasi-1D-CZSNTs above proposed follows from ESR-studies in rather perfect quasi-1D-CZSNTs, produced by HEIBM  of diamond single crysatals, by which the appearance of  Peierls transition and neutral paramagnetic (with spin S = 1/2) SSH-$\pi$-soliton formation were established \cite{Ertchak}, \cite{Ertchak_Stelmakh}. By the way, it means, that perfect  quasi-1D CZSNTs, characterized by $(m,0)$ indices, will have bandgap like to  classical semiconductors  at any $m$, including the case $m = 3q, q \in N$, for which the 2D-theory, existing at present, predicts the metallic properties.  It follows also from Slater principle the following.  Longitudinal
and transverse optical phonon graphite-like  G-modes \cite{Reich}, undergoing in 2D-SWNTs splitting into $G^+$ and $G^-$modes, respectively, \cite{G.Dresselhaus} because of the
curvature effect and
the out-of-plane  $RBM$ \cite{Dresselhaus M.S}, that is all the modes, which  are observed in Raman spectra of  2D-SWNTs, have to disappear  in perfect quasi-1D  CZSNTs. Really, the fact,  that neutral SSH-$\pi$-solitons are responsible for ESR-spectra in quasi-1D  CZSNTs incorporated in diamond matrix is direct confirmation to given conclusion. Neutral (zero charged) SSH-$\pi$-solitons are optically inactive \cite{Heeger_1988}. Consequently all $\pi$-subsystem will be inactive in optical spectra of quasi-1D  CZSNTs. It seems to be one of the most substantial characteristics of 2D-1D transition in physical properties of CZSNTs.

It seems to be interesting to predict, what kind of lines have to appear in Raman spectra instead of $G^+$,  $G^-$ and $RB$-modes by 2D-1D transition.  The model of quasi-1D  CZSNTs  proposed in \cite{Part1}, experimental results on ESR-studies of quasi-1D  CZSNTs and on optical studies of related carbon chain material - carbynes allow to obtain given prediction a priori without detailed analytical calculation. It is sufficient to take into consideration, that SSH-model along with the physical basis of the existence of solitons, polarons, breathers, formed in $\pi$-electronic subsystem ($\pi$-solitons, $\pi$-polarons, $\pi$-breathers) contains in implicit form  also the basis for the existence of similar quasiparticles in  $\sigma$-electronic subsystem, that is  SSH-model can be developed. It was done in \cite{Yearchuck_PL} and in \cite{D_Y}. 

 \subsection{$\sigma$-Quasiparticles in SSH-Model}

 The origin of quasiparticles' formation  in  $\sigma$-electronic subsystem is the same two-fold degeneration of ground state  of the whole electronic system, energy of which in ground state has in correspondence with results in the first subsection the form of Coleman-Weinberg potential with two minima  at the values of dimerization coordinate $u_0$ and $-u_0$ for both kinds of quasiparticles in SSH-model. Really the appearance of $u_0 \neq 0$ and $-u_0 \neq 0$ indicates on the alternation in interatomic distance. It means, that simultaneously with $\pi$-subsystem, $\sigma$-subsystem will also be  dimerized. 

The shapes, for instance, of $\pi$-solitons and
$\sigma$-solitons can be given by the expression with the same mathematical form
\begin{equation}
\label{Eq16}
|\phi(n)|^2 = \frac{1}{\xi_{\pi(\sigma)}} sech^{2}[\frac{(n-n_0)a}{\xi_{\pi(\sigma)}} - v_{\pi(\sigma)} t] \cos \frac{n \pi}{2},
\end{equation} 
where $n, n_0$ are variable and fixed numbers of $CH$-unit in $CH$-chain, $a$ is $C-C$ interatomic spacing projection on chain direction, $v_{\pi(\sigma)}$ is $\pi$($\sigma$)-soliton velocity, $t$ is time, $\xi_{\pi(\sigma)}$ is $\pi$($\sigma$) coherence length. It is seen, that $\pi$-solitons and $\sigma$-solitons differ in fact the only by numerical value of coherence length. Given difference can be evaluated even without numerical calculation of the relation, which determines the shift of ground state energy of extended system by presence of localized perturbation. Actually it is sufficient  to take into account the known value of $\xi_\pi$ and relationships \cite{Lifshitz}
\begin{equation}
\label{Eq17}
\xi_{0\pi} = \frac{\hbar v_F}{\Delta_{0\pi}}, \xi_{0\sigma} = \frac{\hbar v_F}{\Delta_{0\sigma}},
\end{equation}
where $\Delta_{0\sigma}$, $\Delta_{0\pi}$ are $\sigma-$ and $\pi$-bandgap values at $T = 0 K$, $v_F$ is Fermi velocity.
Theoretical value $\xi_\pi$ in t-PA is $7a$, and it is low boundary in the range 
$7a - 11a$, obtained for $\xi_\pi$ from experiments \cite{Heeger_1988}. Taking into account the relationships (\ref{Eq17}),  using the value $\frac{\Delta_{\sigma}}{\Delta_{\pi}}\approx 8.8$, which was evaluated from t-PA band structure calculation in \cite{Grant}, and mean experimental value of coherence length $\overline{\xi_{\pi}} = 9a$ we obtain the value $\overline{\xi_{\sigma}} \approx 0.125 nm$. It means, that halfwidth of space region, occupied by $\sigma$-soliton in t-PA is $\approx 0,25 nm$, that is SSH-$\sigma$-solitons are much more localized in comparison with SSH-$\pi$-solitons. Similar conclusion takes place for SSH-$\sigma$-polarons representing itself the soliton-antisoliton pair.   SSH-$\sigma$-polarons have recently been experimentally detected in the work \cite{Yearchuck_PL}, where the formation of polaron lattice (PL) was established. It was found, that  two components of each elementary unit, that is, of each polaron, possess by two equal in values electical own dipole  moments, proportional to spin, which was called electical spin moments (ESM), with opposite directions. It was shown, that experimental results agree well with  PL-formation, which means in fact  
 the formation of antiferroelectrically ordered lattice of quasiparticles. Given lattice consists of 2 sublattices, corresponding to soliton and antisoliton components of polaron. Corresponding chain state is optically active and it is characterized by  the set of lines in infrared (IR) spectra, which were assigned with new optical phenomenon - antiferroelectric spin wave resonance (AFESWR). Let us remember that carbynes are organic quasi-one-dimensional conductors with the simplest, consisting the only of the carbon atoms, chain structure. At the same time the presence of two electronic  $\pi_x$ and $\pi_y$-subsystems, which are "hung" on a single $\sigma$-subsystem means, that the ground electronic state is similar to two-dimensional Coleman-Weinberg potential with four minima at the values of dimerization coordinate $u_0$ and $ -u_0$. In other words, ground electronic state in carbynes is four-fold degenerate, which leads to a substantially more rich spectrum of possible quasiparticles, discussed in \cite{Rice} and in \cite{Yearchuck_ArXiv}.

Taking into accout that $\sigma$-subsystems are very similar in carbynes and in t-PA and consequently in an arbitrary chain of n-chained  quasi-1D  CZSNTs, we can evaluate the numerical characteristics of AFESWR in in t-PA and in quasi-1D  CZSNTs. Really,
central mode, that is  AFR  mode has  the frequency value $\nu^\sigma_p(C)$ in carbyne sample  equaled to 477 $cm^{-1}$, the splitting parameter in IR detected AFESWR spectra was equal to 150 $cm^{-1}$. 
Given values of $\nu^\sigma_p(C)$ and AFESWR-splitting parameter   in carbynes allow to estimate the range for expected values of analogous parameters in t-PA and in quasi-1D  CZSNTs in the following way. AFESWR-splitting parameter is determined by exchange integrals in $\sigma$-electronic subsystem \cite{Yearchuck_PL}, which seems to be practically the same in carbynes and in t-PA and in quasi-1D  CZSNTs, since the role of quite different $\pi$-subsystems in carbynes and in t-PA and quasi-1D CZSNTs can be neglected to a first approximation. Consequently, the value of AFESWR-splitting parameter in quasi-1D  CZSNTs and in t-PA has to be close to 150 $cm^{-1}$ and to 300 $cm^{-1}$ by IR- and RS-AFESWR-detection correspondingly \cite{Yearchuck_Doklady}. The frequencies $\nu^\sigma_p(C)$ and $\nu^\sigma_p(t-PA)$, $\nu^\sigma_p(NT)$ of main AFESWR-mode in SSH-$\sigma$-polaron lattice in carbyne and in t-PA and in quasi-1D  CZSNTs
 depend on intracrystalline field \cite{Yearchuck_PL}, that means, that their values will be different. However the values of $\nu^\sigma_p(t-PA)$, $\nu^\sigma_p(NT)$ can be evaluated, if to take into account  the known relation for the vibration frequencies of similar centers,
the fact of $\sigma$-polaron and $\pi$-soliton lattice formation in carbynes, leading to change in effective masses \cite{Yearchuck_ArXiv}, presence of two  $\pi$-subsystems  in carbynes,  difference of coherence lengths in accordance with  (\ref{Eq2}) of $\sigma$-solitons ($\xi_\sigma$) and $\pi$-solitons ($\xi_\pi$), band structure  data  for t-PA  \cite{Grant} and carbynes \cite{Leleiter}.

We have obtained the following frequency ranges for IR SSH-$\sigma$-polaron lines  in t-PA and in quasi-1D CZSNTs
$\nu^\sigma_p(t-PA) \in$ (386.7, 603) $cm^{-1}$ and $\nu^\sigma_p(NT) \in $ (402.5, 627.6) $cm^{-1}$.  Known IR-mode with the frequency near 540 $cm^{-1}$, in t-PA \cite{Heeger_1988} gets to interval (386.7, 603) $cm^{-1}$ and can represent itself AFR  mode in $\sigma$-polaron lattice, that is, there is alternative interpretation of given IR-mode, ascribing earlier to Goldstone SSH-$\pi$-soliton vibration mode \cite{Heeger_1988}. 
It follows from results of \cite{Eklund}, that the same  spectral interval with slightly different right-hand value, equaled to  673.7 $cm^{-1}$, is the evaluation for the frequency of $\sigma$-polaron main AFESWR-modes in t-PA and   in quasi-1D  CZSNTs, which are RS-active.  The calculation  in \cite{Eklund} does not take into consideration the soliton and polaron formation. However $\sigma$-polaron  formation does not violate the symmetry of task, that allows to conclude, that the same asymmetry value will be retained  in IR  and RS spectral distributions.

Therefore the qualitative semiclassical consideration predicts a priori, that Raman spectrum of quasi-1D CZSNTs will consist of the only  one line, representing itself AFR-mode of 
  $\sigma$-polaron lattice with peak frequency position in the range (386.7, 673.7) $cm^{-1}$. It can be splitted into series of AFESWR-modes with average splitting parameter near 300 $cm^{-1}$. The possibility of AFESWR-splitting depend on experimental geometry conditions (that is, whether is an antiferroelectric spin wave modes' excitation allowed by the experimental geometry or not).  

It will be shown further, that the prediction has to be completed, if to take into account the quantum nature of EM-field.

Experimental confirmation for conclusion on 2D-1D transition and its main physical properties predicted follows
from ESR-studies, performed earlier in rather perfect CZSNTs, produced by
HEIBM of diamond single
crysatals \cite{Erchak}, \cite{Ertchak}, \cite{Ertchak_Stelmakh}, and from Raman scattering data, presented in Section II, obtained on the same samples.  Let us reproduce the formulation of the model of quasi-1D CZSNTs, given in \cite{Part1}. Quasi-1D CZSNT represents itself autonomous dynamical system with discrete circular symmetry consisting of finite number $n \in N$ of carbon backbones of t-PA chains, which are placed periodically along transverse angle coordinate. Longitudinal axes $\{x_i\}, i = \overline{1,n}$, of individual chains can be directed both along element of cylinder and  along generatrix  of any other smooth figure with axial symmetry. It is taken into account, that Slater principle like to SSH-model for t-PA allows to consider to be active the only degree of freedom along axes $\{x_i\}, i = \overline{1,n}$, of individual chains or, in other words, along single hypercomplex axis in a hypercomplex number language of the task formulation.  It is the reason, that the adjacent chains, which represent themselves a mirror of each other in real structure, becomes to be equivalent in the model, that is, they will be indistinguishable, since the only one  degree of freedom - dimerization coordinate $u_m$ of the $m$-th $C$-atom, 
$m = \overline{1,N}$ along chain molecular-symmetry axis $x$ is substantial for determination of main physical properties in the frames of the model proposed.

 \section{Discussion}

 \subsection{Summarization of Results on HEIBM-Method of Nanotubes Production}

Let us give short review concerning HEIBM-method of production of incorporated carbon NTs and to represent the exact experimental proof of their real formation in diamond matrix. 

The first report on the discovery of new carbon phase - carbon nanotubes, incorporated in diamond matrix,  is related to 1990, and it  was made during 1990 IBMM-Conference, Knoxwille, USA. Similar report  was also represented at E-MRS 1990 Fall Meeting, Strasbourg, France, \cite{Efimov}, that is, the conclusion on new carbon phase formation became to be known substantially prior of now well known  japan discovery of free standing  NTs, reported in \cite{Iijima}, which is related to 1991. Basis experimental method was ESR.
 It was found, that central place takes in ESR spectra of diamond single crystals, modified by high energy ion implantation, the single intensive line with very unusual radiospectroskopic properties. It was anisotropic, however the anisotropy was  weak in comparison with the anisotropy of point centers in diamond. Especially interesting, that along with $g$-value  the linewidth $\Delta{H^{pp}}$ was also found to be tensor quantitity, at that $g$-tensor and $\Delta{H^{pp}}$-tensor were characterized by the same axial symmetry group \cite{Erchak}. Moreover the only one eqivalent configuration in diamond lattice was presented with $g_{\mid\mid}$, $\Delta{H^{pp}_{\mid\mid}}$ principal directions of axial $g$- and $\Delta{H^{pp}}$-tensors, which were strictly coinciding with ion beam direction, at that $g_{\mid\mid}$ is minimal $g$-value,  $\Delta{H^{pp}}$ is maximal $\Delta{H^{pp}}$-value in $g$ and $\Delta{H^{pp}}$ angular dependences.  Other lattice equivalent configurations were absent. The kind of symmetry group, corresponding to  $g$-tensor and $\Delta{H^{pp}}$-tensor symmetry was strongly dependent on  the choice of ion beam direction relatively the crystallographic axes of diamond lattice.  It was shown for the first time in radiospectroscopy, that $g$-tensor and $\Delta{H^{pp}}$-tensor symmetry corresponding to the most intensive line in the spectrum observed are the mapping of the symmetry of lengthy objects with macrosizes along implantation direction \cite{Erchak}. Detailed studies of ESR spectral  angular dependences in appropriate crystallographic planes allowed to establish, that the  structures formed by $\langle 111 \rangle$-HEIBM in diamond have tracklike cylindrical shape, which are lengthy strictly in $\langle 111 \rangle$-direction, with the size varying from several mkm to several tens of mkm, depending on ion energy used. In other words, the first studies in 1990 allowed to establish the formation of carbon nanotubes of cylindical symmetry shape incorporated in diamond matrix. Simultaneously the formation of quite different nanotubes was found. They represent themselves crimped  cylinder with crimping corresponding to four-petal structure in the cross section.  They are produced by the implantation direction, coinciding with $\langle 100 \rangle$ axis of diamond lattice \cite{Erchak}. 

 The next step was the additional confirmation of the structure of cylindical symmetry shape to be really $\langle 111 \rangle$-incorporated nanotubes with rolled up graphene sheet in zigzag shaped configuration (but not for example cylindical rod). It was done by the study of radiospectroscopic properties of spin curriers in given tubes. It was found, that the system of paramagnetic centers (PC), which are responsible for appearance of strong single line absorption above described is non-Blochian system. A number of distinctive peculiarities have been observed for the first time in radiospectroscopy at all. The main ones among them are the following.

1.It was found, that the shape of resonance lines cannot be described by Lorentzian, Gaussian, or by their convolution. The shape is characterized by essentially more slow decrease of the absorption intensity on the wings in comparison with known shape functions. It was called super-Lorentzian. It was proved, that super-Lorentzian is genuine lineshape of resonance absorption (that is, it does not represent the superposition of  Lorentzians with different linewidths) \cite{Ertchak}.

2.It was established, that PC-system is nonsaturating with superlinear absorption kinetics (that is, the dependence of resonance signal amplitude on the amplitude of  magnetic component of microwave field is  superlinear). Especially interesting, that superlinear absorption kinetics  has been observed by both the registration of resonance signal in phase with high-frequency modulation  and in  quadrature  with high-frequency modulation \cite{Erchak}, \cite{Ertchak}.

3.It was found unusual for nonsaturating resonance systems dependence of resonance signal amplitude on the value of modulation frequency, which was characterized by substantial increase (instead decrease) of resonance signal amplitude with modulation frequency increasing \cite{Erchak}, \cite{Ertchak}.

4.It was established the effect of  appearance  of "phase angle", characterizing the absorption process. It consist in that, that although absorption kinetics is nonsaturating, the maximum of absorption is achieved not strictly in phase with modulation field, that is at modulation phase value, which is nonequal to zero. "Phase angle" is anisotropic in general case and maximum of its value was found to be equal $20\textdegree$ \cite{Ertchak}. 

The analysis of given peculiarities \cite{Ertchak}, \cite{Ertchak_Stelmakh} has led to  conclusion, that PC, which are responsible for strong non-Blochian absorption, are mobile and are characterized  by two times of spin-lattice relaxation, by $T_1$, which is comparable with relaxation time  of usual point PC in diamond single crystals $( 10^{-3} - 10^{-4} s)$ and by very short time $\tau$ (upto $10^{-13} s$)  of conversion of the energy of the spin system into mechanical kinetic energy 
of $PC$ motion, which can be considered to be consequence of fundamental in magnetic resonance phenomenon of gyromagnetic coupling between magnetic and mechanical moments. Therefore $\tau$-process is very fast process in comparison with any process, realized by means of pure  magnetic interaction, the  relaxation times of which  are not shorter, than $10^{-10} s$ \cite{Abragam}.  
The possibility of given conversion  can be realized naturally, if the energy of the motion activation is very small, it  is typical for mobile topological solitons like them ones, identified in $t$-PA. So the conclusion, that the strong non-Blochian absorption is determined by mobile topological solitons was obtained by natural way. All subsecuent studies have confirmed given  conclusion. It was found, that there is numerical coincidence of the characteristics of SSH topological $\pi$-solitons in t-PA and in  $\langle 111 \rangle$ incorporated NTs. So, the values of g-tensor components in $Cu$-implanted sample are $g_1$ = 2.00255 (it is minimal $g$-value and it is $g_{\mid\mid}$ principal direction), $g_2$ = $g_3$ = $g_\perp$ = 2.00273, the accuracy of relative g-value measurements is $\pm 0.00002$ \cite{Erchak}. $g$-value of paramagnetic $\pi$-solitons in trans-polyacetylene, equaled to 2.00263 \cite{Goldberg}, gets in the middle of given rather narrow interval of $g$-value variation of PC in ion produced NT. Although anisotropy of paramagnetic $\pi$-solitons in $t$-PA, which are also considered to be mobile PC, mapping the distribution of $\pi$-electron density along whole individual $t$-PA chain, is not resolved by ESR measurements directly (which in fact is the indication, that chemically
produced $t$-PA is less perfect in comparison with NTs in diamond matrix), there are indirect evidences on axial symmetry of ESR absorption spectra in $t$-PA too, \cite{Kahol}, \cite{Kuroda}. Consequently, the value 2.00263 is mean value and it coincides with accuracy 0.00002 with mean value of aforecited principal $g$-tensor values of PC in NTs. Given coincidence becomes to be understandable now, if to take into account the results of \cite{Part1}, where is shown, that model of CZSNTs, produced by HEIBM of diamond single crystals is very similar to SSH-model of t-PA. The main difference of both the models consist in the following. Carbon chain backbone of t-PA single chain  can be considered to be 1D-object in the space $3D \otimes Z_1$, carbon chain backbone of single CZSNT can be considered to be $1D$-object in $3D \otimes Z_n$  space. Here $3D$ is real Euclidian space, $Z_n$ is hypercomplex commutative ring, which is direct sum of $n$ complex spaces $C$. 
\begin{equation}
\label{Eq18}
Z_n = C \oplus {C} \oplus{...}\oplus {C},
\end{equation}
$Z_1 = C$.
 We see, that average ESR properties of $1D$-objects in  $3D \otimes Z_1$ space and in $3D \otimes Z_n$  space are identical with high precision. 

On the other hand we see, that it was possible to determine anisotropy of $g$-values with very high precision, that indicates on the very perfect and homogeneous axially symmetric NTs with symmetry axis strictly along ion beam direction and to establish the origin of PC  in ion produced NTs to be paramagnetic topological $\pi$-solitons (SSH-solitons), considered to be mobile PC, mapping the distribution of $\pi$-
electron density along whole individual NT \cite{Ertchak}, \cite{Ertchak_Stelmakh}.  
   Thereby immediately from ESR studies was found, that structural element of new carbon phase with cylindrical symmetry produced in diamond by $\langle{111}\rangle$ high energy implantation is t-PA chain backbone.  At the same time t-PA chain backbone is basic element for  graphene, graphite, NTs formation and also for the carbyne formation. Carbynes were studied by ESR, they characterized by quite different ESR-spectra \cite{Ertchak_J_Physics_Condensed_Matter}.  The possibility of formation of cylindrical graphite rod  is also not corresponding, since  graphite is characterized by the other ESR spectra. So it was concluded unambiguously, that the spectra observed by $\langle{111}\rangle$ high energy implantation are corresponding the only to graphene sheet, which is rolled up in that way, in order to t-PA chain backbones were directed along $\langle{111}\rangle$-axis of diamond lattice, coinciding with ion beam direction, that results in CZSNTs production. 

Therefore the task of identification of electronic and geometrical structure of CZSNTs was solved by  ESR spectroscopy methods, which are the most powerfull methods [it seems to be commonly accepted opinion] in the structural studies of nanosize objects. Thousands spectra, including the spectra reported in literature for natural diamonds single crystals both in starting materials and irradiated with electrons, neutrons, ions and all known ESR data in related carbon materials - polycrystalline diamond films, fullerenes, coals and graphite were studied and analyzed.   

It is substantial, that the CZSNTs are not random in orientation. They are produced strictly in ion beam direction and they have usual length by $\sim 1 Mev$ per nucleon ion energy, it is proved in details in the works above cited. Naturally random processes are also presenting by HEIBM. The knocked diamond atoms produce point defects in lateral to ion track directions, their concentration distribution in the sample represents itself the overlap of random distributions along directions, which are perpendicular to each single ion track axis. They were also detected and studied by ESR, see, for instance \cite{Erchak_Gelfand}. 
The main difference of our results on NTs from well known ones is based on other dimensionality of CZSNTs studied. We are dealing with quasi-one-dimensional CZSNTs which have to possess and really possess by quite other physical properties including optical properties in comparison with two-dimensional CZSNTs, which have naturally graphite-like optical spectra.
In addition we see also, that the spectra observed by Raman spectroscopy in quasi-one-dimensional CZSNTs can be predicted qualitatively, if to take into consideration the only ESR data (optical data in carbynes were used additionally for quantitative evaluation of optical characteristics, that is AFESWR characteristics).

It was above indicated, that free standing nanotubes were considered theoretically  the only  to be $2D$-strutures.   At the same time there are along with experimental ESR results on SWCNTs incorporated in diamond matrix above summarized, the  experimental ESR and Raman scattering results on narrow free standing tubes with diameter $ < 1 nm$, indicating , that they  cannot be considered strongly speaking to be  $2D$-systems, they are quasi-$1D$ systems. Really, in \cite{Wang} is reported on the development of  SWCNTs of 0.4 $nm$ diameter – the smallest so
far - inside the nanochannels of porous zeolite $AlPO4-5$  single crystals and the authors are considered 0.4 $nm$ diameter NTs to
be ideal one-dimensional quantum hollow wires. The electronic and magnetic properties
of these ultra-small NTs seem to be drastically different from those of large sized
NTs.  Really given tubes were shown to exhibit unusual
novel phenomena like diamagnetism and superconductivity at low temperatures \cite{Tang} in
addition to several other optical properties and doping induced effects.
It seems to be the consequence of one-dimensionality of given NTs.
The direct confirmation of given conclusion is recent ESR-studies in \cite{Rao}, where   
ESR
measurements on ultra-small single walled carbon nanotubes embedded in a SAPO 5
zeolite matrix with a main point of attention to potentially occurring CESR (ESR on the electrons in c-band) signals.
Instead, only one paramagnetic signal was observed of symmetric shape at $g = 2.0025$ on the CNTs in zeolite cages, the ESR
signal exhibits a predominantly Gaussian character at $8.9 GHz (4.2 K)$, with a $\Delta{B_{pp}} = 8.3 G$ and $g = 2.00251$. For the free standing SWCNTs, the ESR signal takes a more
Lorentzian shape at $8.9 GHz (4.2 K)$ with $\Delta{B^{pp}} = 5 G$ and with $g = 2.00254$. So,
from the above observed distinct changes in the ESR spectral properties of NTs with
and without cages, it was inferred in \cite{Rao}, that the ESR signal indeed stems from the carbon NTs
presenting inside the nanochannels of the zeolite. The  observed $g$-value (2.00251) is
inconsistent with the earlier measured values $(g = 2.05 - 2.07)$ \cite{Chauvet}  of CESR on non
embedded single- and multiwalled NTs of larger diameter.
There was observed in \cite{Rao} a dependence of peak-to-peak ESR signal
width in confined SWCNT@SAPO 5 at 70 $K$  on microwave frequecy $f$
suggesting a  inhomogeneous (Gaussian) contribution to the line broadening.
 Least–square linear fit of $\Delta{B^{pp}}$  is $\Delta{B^{pp}(G)}$ = $7.7 + 0.018 f (GHz)$. Inhomogeneous  contribution to the line broadening is clear evidence, that the so called ultra-small carbon nanotubes, studied in \cite {Rao} are not perfect. Nevetheless they can be really considered to be quasi-one-dimensional objects. It means, that like to $t$-PA,  Peierls transition has to take place. Direct indication to given conclusion is $g$-value, which corresponds the only to deep centers in bandgap in any carbon materials \cite{Ertchak}, \cite{Stelmach}. On the one hand it is indicating of the appearance of bandgap in starting metal tubes to be consequence of Peierls metal-semiconductor transition. On the other hand $g = 2.00254$ is  coinciding {in limits of accuracy of measurements) with principal $g$-value   $g_1$ = 2.00255 of axial $g$-tensor of  SSH $\pi$- -solitons in CZSNTs, incorporated in diamond matrix  by $\langle 111 \rangle$ diamond HEIBM and it is near to the $g$-value of paramagnetic SSH $\pi$-solitons
in trans-polyacetylene, equaled to 2.00263 \cite{Goldberg}, that indicates on the similarity of the structure of tubes incorporated in diamond and embedded in zeolite. At the same time angular dependences of $g$-value and $\Delta{B^{pp}}$ were not observed  in \cite {Rao}. It seems to be additional argument to conclusion on insufficient homogeneity of NTs in cited work. Really they have different chiralities (authors report on  the presence of  three chiralities (5,0) (4,2) and (3,3)). On the other hand the observation of spin-glass type of temperature dependence of $\Delta{B^{pp}}$ seems to be the indication, that the  SSH $\pi$-solitons are pinned and have random  distribution along tube axis. For comparison, the temperature dependence of $\Delta{H^{pp}}$ in NTs, produced by HEIBM is quite different and it indicates, that   $\Delta{H^{pp}}$ is not determined by unresolved g-anisotropy or unresolved hyperfine interaction \cite{Erchak}. It is interesting, that pinned solitons can be considered to be similar to chemical radicals, possessing by dangling bonds, which to some extent is in agreement with interpretation of PC-nature to be dangling bond defects  in \cite{Rao}, however they are not usual dangling bond point defects, but they are topological defects. Random  distribution along tube axis means in turn, that individual tube with defined chirality is also not sufficiently perfect to observe optical quantum coherent effects. It means, that Slater principle  with one preferential direction will be not applicable and optical spectra, in particular Raman spectra, will be determined the only by local properties of optical centers, that is, they can be similar to some extent to those ones in the tubes of larger diameters. It is substantial, that main Raman modes in the tubes of larger diameters seem to be similar in both perfect and in not sufficiently perfect NTs, since even in perfect NTs in given case according to above consideration there are two preferential direction to be existing and one-dimensional quantum coherent CZSNT-model \cite{Part1} will be not applicable for description of optical properties in given tubes.

 \subsection{Quantum Field Effects in Raman Scattering in Carbon Zigzag-Shaped Nanotubes}

  Semiclassical evaluation, obtained above has led to  conclusion, that $\pi$-subsystem of quasi-1D CZSNTs will be inactive in optical experiments. The comparison with optical properties of related carbon chain material - carbyne - has allowed to predict, that among possible optically active topological defects in  $\sigma$-subsystem of quasi-1D CZSNTs the $\sigma$-polaron is expected to be prevailing. Therefore intensive line $656.8{\pm}0.2$ $cm^{-1}$ in RS-spectrum, presented in Figure 1, can be assigned with AFR mode of $\sigma$-polaron lattice, produced in CZSNTs studied. Further, theoretical model, proposed in \cite{Part1} allows to insist, that the lines, $1215{\pm}1$ $cm^{-1}$, $1779.5{\pm}1$ $cm^{-1}$ and  $2022.3{\pm}0.5$ $cm^{-1}$ represent themselves the revival part \cite{Slepyan_Yerchak} of Rabi wave packet in its frequency representation \cite{Part1}, which is result of strong interaction of external EM-field with $\sigma$-polaron lattice.  

To make given conclusion to be understandable for the readers let us reproduce the main moments, concerning theoretial model above cited.
Each $\sigma$-polaron interacting with external EM-field in accordance with experiment in \cite{Yearchuck_PL} can be approximated like to guantum dots in \cite{Slepyan_Yerchak} by two-level qubit. Then  the Hamiltonian, proposed in the work \cite{Slepyan_Yerchak}
 was generalized.  The insufficient for the model local field term was omitted. (Local field term seems to be playing minor role by description of $\sigma$-polarons in comparison with 	quantum dots, since size of quantum dots is greatly exceeding the size of $\sigma$-polarons).  The apparatus of hypercomplex $n$-numbers was used.  Hypercomplex $n$-numbers are defined to be elements of commutative ring, given by (\ref{Eq18})
that is, it is direct sum of $n$ fields of complex numbers $C$, $n\in N$. It means that any hypercomplex $n$-number $z$ is $n$-dimensional quantity with the components $k_\alpha \in C$, $\alpha = \overline{0,n-1}$, that is in row matrix form $z$ is
\begin{equation}
\label{Eq19}
z = [k_0, k_1, k_2, ..., k_{n-1}],
\end{equation}
 it can be represented also in the form
\begin{equation}\label{Eq20}
z = \sum_{\alpha = 0}^{n-1}k_\alpha\pi_\alpha,
\end{equation}
where $\pi_\alpha$ are basis elements of $Z_n$ (and simultaneously  basis elements of the linear space of $n$-dimensional lines and  $n$-dimensional row matrix). They are
\begin{equation}
\label{Eq21}
\begin{split}
&\pi_0 = [1,0, ...,0,0],  \pi_1 =[0,1, ...,0,0],\\
&..., \pi_{n-1} = [0,0, ...,0,1].
\end{split}
\end{equation}
Basis elements $\pi_\alpha$ possess by projection properties
\begin{equation}
\label{Eq22}
\pi_\alpha\pi_\alpha = \pi_\alpha\delta_{\alpha\beta}, \sum_{\alpha = 0}^{n-1}\pi_\alpha = 1, z \pi_\alpha = k_\alpha\pi_\alpha
\end{equation}
In other words, the set of $k_\alpha \in C, \alpha = \overline{0, n-1}$ is the set of eigenvalues of hypercomplex $n$-number $z \in Z_n$, the set  of $\{\pi_\alpha\}$, $\alpha = \overline{0, n-1}$ is eigenbasis of $Z_n$-algebra.
 
Then the QED-Hamiltonian, considered to be hypercomplex operator $n$-number, for $\sigma$-polaron system of interacting with EM-field ZSCNTs, consisting of $n$ backbones of $t$-PA chains, which are connected between themselves in that way, in order to produce rolled up graphene sheet in matrix representation is \cite{Part1}
\begin{equation} 
\label{Eq23}
[\hat{\mathcal{H}}] = [\hat{\mathcal{H}}_{\sigma}] + [\hat{\mathcal{H}}_F] + [\hat{\mathcal{H}}_{\sigma F}] + [\hat{\mathcal{H}}_{\sigma \sigma}].
\end{equation}
The rotating wave approximation and  the single-mode approximation of EM-field are used. All the components in (\ref{Eq23}) are considered to be hypercomplex operator $n$-numbers and they are the following. $[\hat{\mathcal{H}}_{\sigma}]$ represents the operator of the energy of $\sigma$-polaron subsystem  in the absence of interaction between $\sigma$-polarons  themselves and with EM-field. It is
\begin{equation} 
\label{Eq24}
[\hat{\mathcal{H}}_{\sigma}] = (\hbar \omega _0/2) \sum_{j = 0}^{n-1}\sum_m {\hat {\sigma}^z_{mj}}[e_1]^j,
\end{equation}
  where $\hat {\sigma}^z_{mj} = \left|a_{mj}\right\rangle  \left\langle a_{mj} \right|-\left|b_{mj}\right\rangle  \left\langle b_{mj} \right|$ is $z$-transition operator between the ground and excited states of $m$-th $\sigma$-polaron in $j$-th chain.  In other words, transition operator technique is used, which was developed in \cite{Yearchuck_Yerchak_Dovlatova}. The second term 
\begin{equation} 
\label{Eq25}
[\hat {\mathcal{H}}_F] = \hbar \omega \sum_{j = 0}^{n-1}\hat {a}^+\hat {a}[e_1]^j 
\end{equation}
is the Hamiltonian of the free EM-field,  represented in the form of 
hypercomplex operator $n$-number. Describtion of the interaction of $\sigma$-polaron sybsystem with EM-field is given by
the component of the Hamiltonian (\ref{Eq23})
\begin{equation}
\label{Eq26}
[\hat {\mathcal{H}}_{\sigma F}] =\hbar g \sum_{j = 0}^{n-1}\sum\limits_m {(\hat {\sigma }_{mj}^+\hat {a}e^{ikma} + \hat {\sigma }_{mj}^-\hat {a}^+e^{-ikma})}[e_1]^j ,
\end{equation}
where $g$  is the interaction constant.
The term $[\hat{\mathcal{H}}_{\sigma \sigma}]$ characterizes intrachain and inter-
chain polaron-polaron interaction. It is given by the relation
\begin{equation}
\begin{split}
\label{Eq27}
&[\hat{\mathcal{H}}_{\sigma\sigma}] =
-\hbar \sum_{l = 0}^{n-1}\sum_{j = 0}^{n-1}\xi^{(1)}_{|l-j|}[e_1]^l\sum\limits_m \left|a_{mj} \right\rangle \left\langle a_{m+1,j}\right|[e_1]^j \\
&-\hbar\sum_{l = 0}^{n-1}\sum_{j = 0}^{n-1}\xi^{(1)}_{|l-j|}[e_1]^l\sum\limits_m \left|a_{mj} \right\rangle \left\langle a_{m-1,j}  \right| [e_1]^j\\
&-\hbar\sum_{l=0}^{n-1}\sum_{j = 0}^{n-1}\xi^{(2)}_{|l-j|}[e_1]^l\sum\limits_m  \left| b_{mj} \right\rangle \left\langle b_{m+1,j}\right|[e_1]^j\\ &-\hbar\sum_{l=0}^{n-1} \sum_{j = 0}^{n-1}\xi^{(2)}_{|l-j|}[e_1]^l\sum\limits_m\left| b_{mj} \right\rangle \left\langle b_{p-1,j} \right|[e_1]^j,
\end{split}
\end{equation}
where  $\hbar\xi^{(1,2)}_{|l-j|}$ are the energies, characterizing intrachain $(l = j)$ and interchain $(l \neq j)$ polaron-polaron interaction for the excited ($\xi^{(1)}$) and ground ($\xi^{(2)}$) states of $j$-th chain, $\left| b_{mj} \right\rangle, \left|a_{mj}\right\rangle$ are ground and excited states correspondingly of $m$-th $\sigma$-polaron of $j$-th chain. In (\ref{Eq24}) to (\ref{Eq27}) $[e_1]^j$ is j-th power of the circulant matrix $[e_1]$, which is
\begin{equation}
\label{Eq28}
[e_1]=\left[\begin{array} {*{20}c} 0&1&0& ...&0  \\ 0&0&1& ...&0 \\ &...& \\ 0&0& ... &0&1\\1&0&...&0&0 \end{array}\right].
\end{equation}
Hamiltonian in the form like to (\ref{Eq27}) at $n = 1$ is usually used for description of tunneling 
between the states with equal energies, in particular, for tunneling between quantum dot states \cite{Slepyan_Yerchak}. Hamiltonian (\ref{Eq27}) at any $n$ describes actually the connection between pairs of the states, which satisfy the following condition - the first state  in any pair results from the second state (and vice versa)  by time reversal. It is known, that for given states Cooper effect takes place. Therefore,  the application of Hamiltonian like to (\ref{Eq27}) is possible for any pair of the states with equal energy, which are symmetric relatively time reversal.  
By the way, if one
omits last  term in Hamiltonian (\ref{Eq23}), it goes into $n$-chain generalization of
well-known Tavis-Cummings Hamiltonian \cite{Tavis}.
 Then the state vector of the "NT+EM-field" system was represented \cite{Part1} in terms of the eigenstates of isolated polaron and photon number states in the folowing matrix form
\begin{equation}
\begin{split}
\label{Eq29}
&[\left| {\Psi (t)} \right\rangle] = \\ 
&\sum_{j = 0}^{n-1}\{\sum\limits_l \sum\limits_{m} \left(A^j_{m,l}(t) \left|a_{mj},l \right\rangle + B^j_{m,l}(t) \left| b_{mj},l \right\rangle\right) \}[e_1]^j.
\end{split}
\end{equation}
Here, $\left| b_{mj},l \right\rangle = \left| b_{mj} \right\rangle\otimes\left|l \right\rangle$, $\left| a_{mj},l \right\rangle = \left| a_{mj} \right\rangle\otimes\left|l \right\rangle$, where  $\left|l \right\rangle$ is the EM-field  Fock state with $l$  photons, $A^j_{m,l}(t)$, $B^j_{m,l}(t)$ are the unknown probability amplitudes. Let us pay the attention,  that matrix function  $[\left| {\Psi (t)} \right\rangle]$ is single state and the representation in the form of sum over j is like to representation of complex-valued wave function $\Psi (t)$  in the form $ \Psi (t) = \Psi_1 (t) + i \Psi_2 (t)$  of two real-valued functions.
Nonstationary Schr\"odinger equation for hypercomplex matrix function $[\left| {\Psi (t)} \right\rangle]$ in the interaction representation is  
\begin{equation}
\begin{split}
\label{Eq30}
 i\hbar \frac{\partial}{\partial{t}}[\left| {\Psi (t)} \right\rangle] = [\hat{V}(t)][\left| {\Psi (t)} \right\rangle],
\end{split}
\end{equation}
where matrix Hamiltonian of interaction $[\hat{V}(t)]$ is
\begin{equation}
\begin{split}
\label{Eq31}
&[\hat{V}(t)] = \exp(\frac{i}{\hbar}[\hat{\mathcal{H}}_F] t)([\hat{\mathcal{H}}_{\sigma}] + [\hat{\mathcal{H}}_{\sigma F}] + \\
&[\hat{\mathcal{H}}_{\sigma \sigma}])\exp(-\frac{i}{\hbar}[\hat{\mathcal{H}}_F] t)
\end{split}
\end{equation}

It leads to the set of difference-differencial matrix equations, which   in compact form are
\begin{equation}
\begin{split}
\label{Eq32}
&\frac{\partial}{\partial{t}}[\Psi_{m,l}(t)] = \{-\frac{i}{2} \omega_0 [\sigma_z] - \frac{\lambda}{2}[E_2] - \\  
&i g \sqrt{l + 1}[\sigma_z] \exp({i[\sigma_z] (\omega t - kma)})\}\  
[\Psi_{m,l}(t)] + \\  
&i [\xi] ([\Psi_{m-1,l}(t)] + [\Psi_{m+1,l}(t)]),
\end{split}
\end{equation}
where $[\sigma_z]$ is Pauli $z$-matrix, $[E_2]$ is $[2 \times 2]$ unit matrix, $[\xi]$ is block matrix
\begin{equation}
\begin{split}
\label{Eq33}
[\xi] = \frac{1}{2} ([\xi_1] + \xi_2]) \otimes [E_2] + ([\xi_1] - \xi_2]) \otimes [\sigma_z]]
\end{split}
\end{equation}
where  $[\xi_1]$, $[\xi_2]$ are $[n \times n]$ matrices of coefficients, determined by (\ref{Eq27}), that is they are
\begin{equation}
\begin{split}
\label{Eq24a}
[\xi^j_1] = \sum_{l = 0}^{n-1}\xi^{(1)}_{|l-j|}[e_1]^l, [\xi^j_2] = \sum_{l = 0}^{n-1}\xi^{(2)}_{|l-j|}[e_1]^l,
\end{split}
\end{equation}
where it is taken into account, that in view of axial symmetry $[\xi^j_{1,2}]$ do not depend on $j$. 
Consequently, we have $[\xi^j_1] = [\xi_1]$, $[\xi^j_2] = [\xi_2]$.
Let us remark, that  the relaxation processes were taken into consideration in (\ref{Eq32}).
It was done by means of  substituting instead 
real value $\omega_0$ the complex values  with imaginary part $i\lambda$.  It is also suggested, that relaxation time $\tau$ is independent on the chain number and it is determined by the value, which is reciprocal to $\lambda$, that is  $\tau = \frac{2\pi}{\lambda}$.

 The solution of hypercomplex equation (\ref{Eq32}) for the state vector $[\left| {\Psi (t)} \right\rangle]$ was obtained   in continuum limit and it is given by the expression
\begin{equation}
\begin{split}
\label{Eq34}
&[\Phi^l(x,t)] = \\
&\int\limits_{-\infty }^\infty{[\overline{\Phi}^l(h,0)]\exp\{i t([\theta^l(h)] - g \sqrt{l+1}[\chi])\}e^{ihx}}dh,
\end{split}
\end{equation}
where $x$ is  hypercomplex axis $x = [x, x, ..., x]$, $[\Phi^l(x,t)]$ is 
\begin{equation}
\begin{split}
\label{Eq35}
[\Phi^l(x,t)] = exp{\frac{i(\omega_0 t - kx)[\sigma_z]}{2}}\exp{\frac{\lambda t}{2}}[\Psi^l(x,t)], 
\end{split}
\end{equation}
 In its turn $[\Psi^l(x,t)]$ is continuous limit of functional block matrix of discrete variable $m$, which is given by 
\begin{equation}
\begin{split}
\label{Eq17e}
[\Psi_{m,l}(t)] = \left[\begin{array} {*{20}c}&[A_{m,l}(t)]\\
&      \\
&[B_{m,l+1}(t)]\end{array}\right],  
\end{split}
\end{equation} 
consisting of two $[n \times n]$ matrices of probability amplitudes
\begin{equation}
\begin{split}
\label{Eq21a}
&[A_{m,l}(t)] = \sum_{j = 0}^{n-1}A^j_{m,l}(t)[e_1]^j, \\
&[B_{m,l+1}(t)] = \sum_{j = 0}^{n-1}B^j_{m,l+1}(t)[e_1]^j,
\end{split}
\end{equation}
which are determined by relationship (\ref{Eq29}).
Further, matrix $[\theta(h)]$ in (\ref{Eq34}) is 
\begin{equation}
\begin{split}
\label{Eq36}
&[\theta(h)] = \frac{1}{2}\{([\theta_1(h)] + [\theta_2(h)]) \otimes [E_2]\} \\
&+ \frac{1}{2}\{([\theta_1(h)] - [\theta_2(h)]) \otimes [\sigma_z]\}, 
\end{split}
\end{equation}
where  $[\theta_1(h)]$ and $[\theta_2(h)]$ are 
\begin{equation}
\begin{split}
\label{Eq37}
&[\theta_1(h)] = [\xi_1]\{2 - a^2(h + \frac{k}{2})^2\}, \\
&[\theta_2(h)] = [\xi_2]\{2 - a^2(h - \frac{k}{2})^2\}
\end{split}
\end{equation}
Here  $[\xi_1]$, $[\xi_2]$ are $[n \times n]$ matrices of coefficients, defined by  (\ref{Eq24a}). Matrix $[\chi]$ in (\ref{Eq34}) is
\begin{equation}
\begin{split}
\label{Eq38}
[\chi] = [E_n] \otimes [\sigma_x]\exp ({-i [\sigma_z] (\omega - \omega_0) t}).
\end{split}
\end{equation}
Matrix elements of $[\Phi^l(x,t)]$ are
\begin{equation}
\begin{split}
\label{Eq39}
&\Phi^l_{qp}(x,t) = \int\limits_{-\infty }^{\infty}\Theta^l_{q}(h,0) \exp{ \frac{-2\pi qpi}{n}} \exp{ihx}\times \\
&\exp{\{i\sum_{j = 0}^{n-1}\exp{\frac{2\pi qji}{n}(\vartheta_j(h) - g\sqrt{l-1}\kappa_j(h))}\}}dh,
\end{split}
\end{equation}
where $\Theta^l_{q}(h,0)$,$\vartheta_j(h)$, $\kappa_j(h)$ are determined by eigenvalues $\textbf{k}_\alpha \in C, \alpha = \overline{0, n-1}$ of $\Phi^l(h,0)$, $\theta(h)$ and $\chi(h)$, which are considered to be $n$-numbers. They are
\begin{equation}
\label{Eq40}
\Theta^l_{q}(h,0) = \frac{1}{n}\textbf{k}_q (\Phi^l(h,0)) = \frac{1}{n}\sum_{j = 0}^{n-1}\Phi_j^l(h,0)\exp{\frac{2\pi q j i}{n}}
\end{equation}
\begin{equation}
\label{Eq41}
\vartheta_j(h) = \frac{1}{n}\textbf{k}_j(\theta(h)) = \frac{1}{n}\sum_{r = 0}^{n-1}\theta_r(h)\exp{\frac{2\pi  jr i}{n}},
\end{equation}
\begin{equation}
\label{Eq42}
\kappa_j(h) = \frac{1}{n}\textbf{k}_j(\chi(h)) = \frac{1}{n}\sum_{r = 0}^{n-1}\chi_j(h)\exp{\frac{2\pi j r i}{n}}.
\end{equation}
Then the hypercomplex solution  can be represented in the form of sum of $n$ solutions for $n$ chains, that is, hypercomplex $n$-number $\Phi^l(x,t) $ is
\begin{equation}
\label{Eq43}
\Phi^l(x,t) = \sum_{q = 0}^{n-1}\tilde{\Phi}^l_q(x,t),
\end{equation}
where the solution for $q$-th chain $\tilde{\Phi}^l_q(x,t)$ is
\begin{equation}
\label{Eq44}
\tilde{\Phi}^l_q(x,t) = \sum_{p = 0}^{n-1}\Phi^l_{qp}(x,t)[e_1]^p,
\end{equation}
in which the  matrix elements $\Phi^l_{qp}(x,t)$ are  determined by (\ref{Eq39}).
The relationship (\ref{Eq44}) by taking into account  (\ref{Eq34}) - (\ref{Eq43}) determines Rabi-wave packet, which propagates along individual chain of zigzag NT.  It is now clear, that subsequent analysis of Rabi-wave packet dynamics for individual NT-component will be the same (by rescaling of parameters), that in \cite{Slepyan_Yerchak}. The parameters can in principle be obtained by detailed comparison of results obtained with experiments on optical absorption, reflection or Raman scattering. It is the subject for futher work. For qualitative analysis we can use arbitrary parameters of the task.
Then 
from  equation (\ref{Eq44}) follows temporal dependence of the integral inversion for  j-th chain in CZSNT  and spectral dependence of the RS-signal amplitude, which explain the appearance of a number lines, which are additional to AFSWR lines in the spectra observed. Really
temporal dependence of the integral inversion gives in implicit form  the way for  comparison of theoretical results with any stationary optical experiments in QD-chain or like them, including quasi-1D CZSNTs, with aforesaid initial state. 
It is sufficient to make a Fourier transform of  given temporal dependence. It will be proportional to signal amplitudes of  infrared (IR) absorbtion, IR-transmittance, IR-reflection or Raman scattering, since they are  determined by population difference. It means, that dynamical nonstationary properties of optical systems can become apparent by conventional stationary registration of the spectra. It is very  similar to the well known in stationary ESR-spectroscopy situation, where the signals of the centers with long relaxation times can be registered in nonstationary regime, which can lead to zeroth absorption (or  appearance of the signals in inverse phase like to  maser effects), that was practically used in very many ESR-studies to unravel complicated overlapped spectra (however without theoretical explanation of given effect). At the same time there is difference of principle between classically considered nonstationary regime with Rabi oscillations by classical EM-field consideration and between QED-consideration. If EM-field is considered to be classical field, then  operating point by the registration of the centers with long relaxation times can move  along Rabi oscillation curve, leading to radical changes in amplitude of the signals, including the appearance of irradiation instead absorption. Naturally in given process the energy of EM-field source is partly  comes back from spin systems in the same quality (in EM-field form) instead its transformation in phonon energy. Given process can be accompanying by enhanced noice level, however without appearance of any additional lines. QED-consideration shows,  that the process  of Rabi waves' formation  is determined essentially by interaction  of optical centers with photon subsystem, at that stationary regime is achieved fastly, giving usual stationary optical signals, however after a time the oscillations emerge again, and given revival part leads to additional lines in high energy part of spectral dependence. Therefore, the  absorption (scattering) process by presence of Rabi waves is always nonequilibrium process in the whole system [EM-field + matter] and it can be realized even at relatively short relaxation times of optical centers, if constant $g$ of the interaction of optical centers with EM-field is large.

Thus, the CZSNT-QED-model predicts the appearance of additional lines, for which Rabi waves are responsible, in stationary optical spectra of quasi-1D CZSNTs. It is understandable, that   Rabi waves can be  registered  in any 1D-systems by stationary optical measurements, if the electron-photon coupling is rather strong.

The identification of the lines $1215{\pm}1$ $cm^{-1}$, $1779.5{\pm}1$ $cm^{-1}$,  $2022.3{\pm}0.5$ $cm^{-1}$ (and corresponding lines in boron ion modified sample) with Fourier-image of revival part of Rabi packet is confirmed by the following. It is well known \cite{Abragam}, that  in the case of point absorbing centers Rabi frequency is linear function of the amplitude of oscillating EM-field. I was experimrntally confirmed in \cite{Yerchak}, \cite{Yerchak_Stelmakh}. Given dependency takes also place for the center of Rabi wave packet, that follows from the analysis of Fourier transform of temporal dependence  of the  integral inversion. Further, it is evident, that amplitude of laser wave, penetrating in IBMR, is lesser by excitation from unimplanted side. It is consequence of some absorption in the unimplanted volume of diamond crystal. We see, that really, the  values 1569 ${\pm}3$ $cm^{-1}$, 1757${\pm}5$ $cm^{-1}$ of two high frequency components are substantially less in given case, than $1779.5{\pm}1$ $cm^{-1}$ and  $2022.3{\pm}0.5$ $cm^{-1}$, observed by excitation from implanted side, Figure 2,  Figure 1. Relative frequency changes are 1,151 and 1.134 ${\pm}0.003$ for the pairs [$2022.3{\pm}0.5$ $cm^{-1}$, 1757${\pm}5$] and [$1779.5{\pm}1$ $cm^{-1}$, 1569 ${\pm}3$] correspondingly and it is seen, that they are close to each other, at that  larger frequency undergoes larger change in correspondence with theoretical analysis.  

 The lines at 354.6,  977.1 (${\pm}1$ $cm^{-1}$ seem to be assigned with two AFESWR 
modes. The substantial decrease of relative intensity of 641.8 ${\pm}1$ $cm^{-1}$ mode in comparison with  $656.8{\pm}0.2$ $cm^{-1}$ mode testifies in favour of given assignment. It is seen, that AFESWR-splitting is rather large and it has thr same order of values with the splitting between two polaron vibronic levels. It means, that linear AFESWR-theory  \cite{Yearchuck_PL},  which predicts   a set of equidistant  AFESWR-modes, arranged the left and the right of central mode, can be used the only to obtain approximately the average value of AFESWR-splitting. Really, in given case AFESWR-modes   are not equidistant, they are shifted on different distances 335.3, 287,2 $cm^{-1}$ from main AFESWR-mode. At the same time average value of AFESWR-splitting is 311.3$cm^{-1}$ and it is close to the value in 300 $cm^{-1}$  expected in accordance with linear AFESWR-theory by taking into account experimental AFESWR-splitting value in carbynes. 

  Therefore, we obtain the direct proof of assigment of the lines 656.8${\pm}0.2$ $cm^{-1}$ and 641.8 ${\pm}1$ $cm^{-1}$ with $\sigma$-polaron lattice.   
 \begin{figure}
\includegraphics[width=0.5\textwidth]{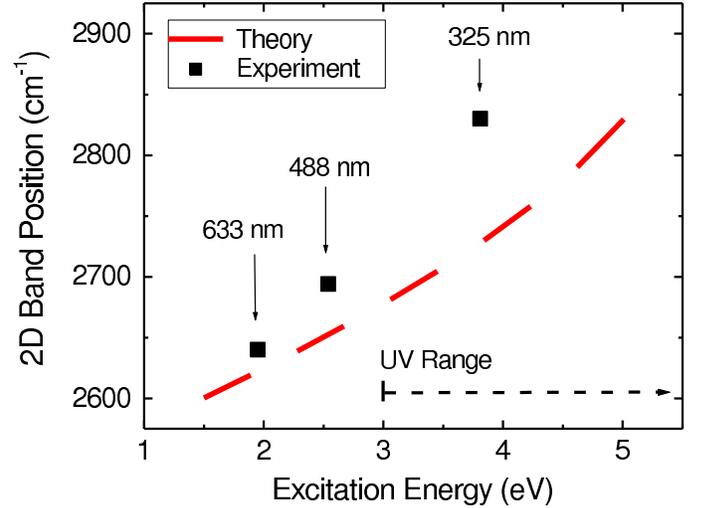}
\caption[Dependence of 2D-band position on excitation energy, presented  in \cite{Calizo} (Figure 5)]
{\label{5} Dependence of 2D-band position in graphene on excitation energy, presented  in \cite{Calizo} (Figure 5)}
\end{figure}
 \begin{figure}
\includegraphics[width=0.5\textwidth]{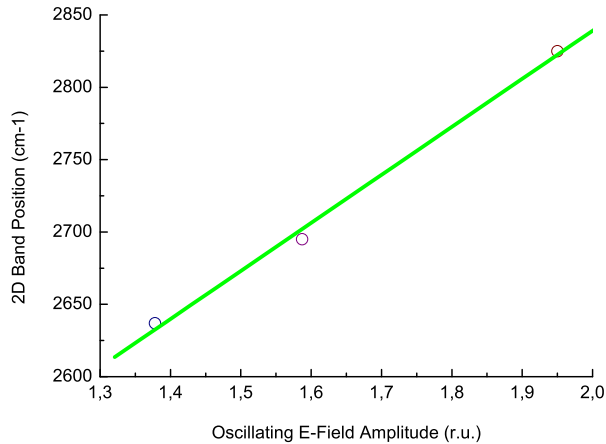}
\caption[Dependence of 2D-band position in graphene on electrical component of oscillating excitation field]
{\label{Graphene} Dependence of 2D-band position in graphene on electrical component of oscillating excitation field}
\end{figure}

To explain  differences  by the change of the direction of the excitation wave propagation, we have to take into consideration the following. The possibility of SWR-excitation in 1D-systems is strongly dependent on  the geometry of experiment, detrmined by  directions of chain axis, vectors of intracrystalline and external magnetic or electric fields (by ferromagnetic or ferroelectric  SWR study correspondingly). It is confirmed by ferromagnetic SWR study in carbynes, \cite{Ertchak_J_Physics_Condensed_Matter}. The axes in CZSNTs are not linear in the end of ion run and they are generatrixes  of the figure of onion-like shape, that provides for necessary geometry for  AFESWR-excitation of $\sigma$-polaron lattice. Moreover the appearance of very broad line, Figure 2, seems to be indication on the excitation  of the Fr\"ohlich movement of $\sigma$-polaron lattice itself. The presence of Fr\"ohlich sliding of $\sigma$-polaron lattice allows to explain qualitatively the appearance of "hysteresis" in spectral dependences relatively changing of the direction of excitation wave propagation into opposite direction. It follows from energy law conservation position. Really, moving $\sigma$-polaron lattice possesses by kinetic energy. It means, that it is required lesser energy value to excite the local polaron vibration mode in correspondence with observation. 

Therefore CZSNTs incorporated in diamond matrix represent themselves the example of the system, which strongly interact with EM-field. Experimental detection of Rabi wave packets confirms on the one hand the theory, developed in \cite{Part1} and it is also the first experimental confirmation of Rabi wave phenomenon, predicted in \cite{Slepyan_Yerchak}, on the other hand.  It means also, that semiclassical description of spectroscopic transitions in CZSNTs and in the systems like them cannot be appropriate. It substantially  raises the practical concernment of QED-theory.   

The requirement of perfect one-dimemensionality is substantial for applicability of Slater principle to CZSNTs, at the same time the Rabi wave phenomenon seems to be general and can be observed in 2D and 3D-systems. For instance, the first-order RS-peak in the vicinity of 1580 $cm^{-1}$ and so called second-order lines  around 2480, 2700 (2D band), 3250 $cm^{-1}$ by $\lambda_{exc}$ = 488 nm were observed in graphene \cite{Calizo}, at that the intensity of second-order 2D-band was found to be substantially exceeding the intensty of the corresponding first-order line. Moreover theoretical dependence of 2D-band position on excitation energy, presented  in \cite{Calizo} by Figure 5, is far from experimental dependence, see Figure 4, which is reproduction of Figure 5 from \cite{Calizo}. At the same time the suggestion, that the second-order lines are spectral mapping  of revival part of Rabi wave packet, corresponding to  the first-order line is agreeing  with experimental data in \cite{Calizo} very well, see Figure 5 in our paper. It wasobtained from experimental data in \cite{Calizo} by choosing of $x$-coordinate instead the energy the electrical component of oscillating excitation field.  It is seen,    that given experimental dependence is linear in full correspondence with expected for the center of Rabi wave packets. It is analogue of well known in radiospectroscopy linear dependece of Rabi oscillation frequency on magnetic  component of oscillating excitation field, see for instance \cite{Yerchak}, \cite{Yerchak_Stelmakh}. 

Therefore given result is experimental evidence of the formation and propagation of Rabi wave packets in graphene, being to be the example of 2D-system.
Similar explanation can be proposed for   transitions in heavily boron-doped
diamond in the region of 1400-2800 $cm^{-1}$ \cite{{Vlasov}}. It means, that really the theory of Rabi wave, presented in \cite{Slepyan_Yerchak} for 1D-systems can be  developed and generalized for 2D- and 3D-systems. Consequently, it seems to be evident, that   Rabi wave packets  can also be identified in free standing NTs by the production technology improvement.  

\section {Conclusions}

 Thus, QED-model for  multichain coupled qubit quasi-1D  system, proposed in \cite{Part1}, is confirmed by Raman scattering studies of carbon zigzag-shaped nanotubes, produced by HEIBM of natural diamond single crystals. New quantum optics phenomenon - Rabi wave packet formation, predicted in \cite{Slepyan_Yerchak},  has been experimentally identified for the first time.  
It is shown, that Raman spectra in  quasi-1D CZSNTs are quite different in comparison with well known Raman spectra in 2D those ones. They characterized by semiclassical consideration by the only  one vibronic mode of Su-Schriffer-Heeger $\sigma$-polaron lattice with peak positions $656.8{\pm}0.2$ $cm^{-1}$, for instance, in copper implanted sample (by excitation from impanted side of the sample) instead of longitudinal
and transverse optical phonon    $G^+$ and $G^-$modes and
the out-of-plane radial breathing mode, which are observed in Raman spectra of  2D single wall nanotubes. It is consequence of 2D - 1D transition in all physical properties of nanotubes. 

It is shown, that  strong electron-photon coupling takes place in quasi-1D CZSNTs by interaction with EM-field and quantum nature of EM-field has to be taken into account. It has been done for the first time in stationary spectroscopy at all. All optical spectra, in particular, Raman spectra are registered by usual stationary measurement technique  in nonequilibrium conditions, which are the consequence of Rabi wave packets' formation. It leads in its turn  to appearance of additional lines with peak positions $1215{\pm}1$ $cm^{-1}$, $1779.5{\pm}1$ $cm^{-1}$ and  $2022.3{\pm}0.5$ $cm^{-1}$ in copper implanted sample and additional lines with peak positions at 1212.3${\pm}1$,  1772.5${\pm}1$, 2011${\pm}0.5$$cm^{-1}$ in boron implanted sample (by excitation from implanted side in both the samples), which are
 corresponding to revival part of inversion dependence of joint EM-field + matter system in the frequency representation.  

The substantial qualitative and quantitative dependence of Raman spectra on changing of the direction of excitation wave propagation into opposite direction has been revealed. Instead of the  RS-lines with peak positions at $656.8{\pm}0.2$ $cm^{-1}$, $1215{\pm}1$ $cm^{-1}$, $1779.5{\pm}1$ $cm^{-1}$ and  $2022.3{\pm}0.5$ $cm^{-1}$, which are obseved by excitation from implated side of the copper imlanted sample, RS-lines with peak positions  at 1328.7 $cm^{-1}$, 354.6, 641.8, 977.1 (${\pm}1$ $cm^{-1}$), 1274.1 ${\pm}2$ $cm^{-1}$  and more weak pronounced lines at 1569 ${\pm}3$ $cm^{-1}$, 1757${\pm}5$ $cm^{-1}$ were observed. Moreover, all the lines listed are superimposed  with very broad (its linewidth  is 1720${\pm}20$ $cm^{-1}$) asymmetric line with peak position at $1160{\pm}10$$cm^{-1}$. RS-lines with peak positions  at 354.6, 641.8, 977.1 (${\pm}1$ $cm^{-1}$) were identified with AFESWR-modes. RS-line with peak position  at 1274.1 ${\pm}2$ $cm^{-1}$  and the lines at 1569 ${\pm}3$ $cm^{-1}$, 1757${\pm}5$ $cm^{-1}$ were reffered  to revival part of inversion dependence of joint EM-field + matter system in the frequency representation. More low frequency values in comparison with those ones observed by excitation from implanted side of the sample are connected with decrease of exciting light intensity to be the consequence of additional absorption in unimplanted region of the sample. RS-line with peak position  at 1328.7 $cm^{-1}$ corresponds to optical phonon in diamond lattice, its value indicates on some softening of optical phonon mode, taking place by copper implantation. It is interesting, that   softening is absent by boron implantation. It seems to be display of size effect relatively effective radius of implated ions. Very broad   asymmetric line with peak position at $1160{\pm}10$ $cm^{-1}$ is attributed to Fr\"ohlich sliding of $\sigma$-polaron lattice.  The appearance of sliding allows to explain qualitatively on the basis of energy conservation law the effect of "hysteresis" in the frequency value of main AFR mode by changing of the direction of excitation wave propagation into opposite direction resulting in its decrease to 641.8 ${\pm}1$ $cm^{-1}$) instead of $656.8{\pm}0.2$ $cm^{-1}$.

\end{document}